\documentclass[12pt]{article}
\usepackage{caption}
\usepackage{authblk}
\usepackage{geometry}
\usepackage{amsmath}
\usepackage{amssymb}
\usepackage{amsthm}
\usepackage{mathrsfs}
\usepackage{customcommands}
\usepackage{bm}
\usepackage{Nettastyle}
\usepackage{dsfont}
\usepackage{comment}
\usepackage[utf8]{inputenc}
\usepackage{calc}
\usepackage{accents}
\usepackage{amsfonts}

\usepackage{braket}
\usepackage[normalem]{ulem}
\usepackage{color}

\usepackage{ulem}
\usepackage{mathtools}
\usepackage{tensor} 
\usepackage{tikz}
\usetikzlibrary{arrows, positioning, quotes, intersections}
\usepackage{subcaption}
\usepackage{graphicx}  

\usepackage{thmtools, thm-restate}

\title{Further Evidence Against a Semiclassical Baby Universe in AdS/CFT}

\author[1]{Netta Engelhardt}
\author[2,3]{and Elliott Gesteau}

\affiliation[1]{Center for Theoretical Physics, Massachusetts Institute of Technology, \\Cambridge, MA 02139, USA}
\affiliation[2]{Division of Physics, Mathematics, and Astronomy, California Institute of Technology,\\
Pasadena, CA 91125, USA}
\affiliation[3]{Kavli Institute for Theoretical Physics, \\Santa Barbara, CA 93106, USA}
\emailAdd{engeln@mit.edu}
\emailAdd{egesteau@caltech.edu}

\abstract{We argue that a large class of asymptotically AdS geometries with a semiclassical baby universe cannot be realized within the AdS/CFT correspondence. This in particular resolves a recent puzzle introduced by Antonini and Rath, in which a single CFT state appeared to simultaneously describe an AdS spacetime with a baby universe and one without.  We construct a low-energy (and low complexity) boundary operator whose expectation values in the descriptions with and without the baby universe cannot match  if the baby universe is semiclassical. This operator conclusively identifies the actual bulk dual: the spacetime \textit{without} a semiclassical baby universe. This result assumes only that AdS/CFT admits an extrapolate dictionary and an asymptotically isometric encoding of the causal wedge into the dual CFT, without which the correspondence may well be vacuous.}

\begin{document}

\maketitle

\section{Introduction}\label{sec:intro}
Recent work on black hole information starting with the calculations of the Page curve~\cite{AEMM, Pen19} has clearly demonstrated that the entanglement wedge of the radiation of an old black hole includes a spatially large compact region in the black hole interior. As the black hole shrinks, the holographic encoding map becomes non-isometric~\cite{AkeEng22}, taking a large Hilbert space in the effective description of the system into a much smaller Hilbert space of the fundamental theory. In the limit that the black hole evaporates fully, the interior resembles a closed baby universe with a singularity that pinches off. The most natural extension of the holographic map into this regime suggests that this singular closed universe has a one-dimensional Hilbert space~\cite{AlmMah19a, PenShe19}. 

This is a radical implication: without some additional degrees of freedom (as suggested in e.g.~\cite{HarUsa25, AbdSte25}), a one-dimensional Hilbert space results in large fluctuations in observables, inconsistent with the local physics experienced in the universe: the baby universe cannot admit a semiclassical description. While it is tempting to dismiss the original argument due to potential failure of the QES prescription~\cite{EngWal14} at the evaporation point, other arguments against semiclassicality of the baby universe, using different tools, have appeared in~\cite{MarMax20, McNVaf20, DonKol24, UsaWan24, UsaZha24, HarUsa25}. For instance, Ref.~\cite{UsaZha24} argued that if a CFT dual to the Euclidean preparation of a state via the Maldacena-Maoz wormhole~\cite{MalMao04} exists in Lorentzian time, the resulting boundaryless geometry cannot be semiclassical. This approach does not suffer from the black hole singularity endpoint issue, but it does involve an unconventional application of the AdS/CFT map to a spacetime that in Lorentzian signature has no AdS boundary. Other arguments have used the QES formula in a smooth baby universe to argue that the Hilbert space of the baby universe must be one-dimensional~\cite{AlmMah19a}. This approach requires an application of the QES formula to a setting in which the latter has not traditionally been applied.

Within the context of the AdS/CFT correspondence, there is a sense, spelled out in a tensor network toy model in~\cite{DonQi20}, in which a closed universe must have a one-dimensional Hilbert space: the CFT Hilbert space is defined via the boundary conditions, and in the absence of a boundary, the CFT Hilbert space must be one-dimensional. This was explored in the general context of the swampland program in \cite{McNVaf20}.

However, an important subtlety must be pointed out at this stage. Indeed, it may be that even though in the \textit{fundamental description} of quantum gravity there are no closed universe degrees of freedom, \textit{effective} semiclassical physics in a closed universe connected component could still somehow be encoded, via some unusual map, into a UV-complete theory that sits at an asymptotic boundary that is disconnected from the closed universe (see e.g.~\cite{DonQi20} for a discussion in a tensor network toy model). This would be in some sense the most extreme possible version of nonlocality, where the physics of a connected component of spacetime would be encoded in a \textit{different} component of spacetime disconnected from the former. However, the holographic map can be indeed be extremely nonlocal:  the entanglement wedge of a boundary region can reach much deeper into the bulk than its causal wedge and even be disconnected from it\footnote{Here we mean as a region rather than as a manifold.}, a feature that was crucial for the holographic realization of the Page curve~\cite{AEMM, Pen19}.  It is therefore not inconceivable in principle that the entanglement wedge of a boundary subregion could reach so far out that it includes a disconnected manifold.

In this article, we show, without making use of toy models or assumptions about AdS/CFT behind horizons, that at least in many cases, including a semiclassical closed universe in the entanglement wedge of a boundary region is too nonlocal, even for the holographic map. In particular, we construct an ${\cal O}(1)$ complexity boundary operator whose expectation value has the power to determine whether a semiclassical baby universe exists or not. Using this operator, we find that in a large class of semiclassical geometries, the most fundamental features of AdS/CFT force the Hilbert space of a closed universe to be one-dimensional \textit{even in the semiclassical description}. Our construction requires nothing more than the AdS/CFT extrapolate dictionary applied to the standard causal wedge of an asymptotically AdS region with ${\cal O}(1)$ energy. This argument uses the baby universe state preparation of Antonini-Rath (AR)~\cite{AntRat24, AntSas23}, which we review in detail Sec.~\ref{sec:ARPuzzle}. We also provide a generalization beyond the AR construction to a class of high energy states at the cost of some additional assumptions about the AdS/CFT correspondence.

Let us briefly explain the gist of the argument. In the AR construction, a two-boundary CFT state $\ket{\Psi}_{AB}$ is prepared in a partially entangled state at some temperature below the Hawking-Page transition\cite{AntSas23}. AR show that the path integral preparation of this state results in two disconnected asymptotically AdS regions with no horizons and a baby universe. AR then argue that this state can also be represented by an ${\cal O}(1)$ number of operators of low conformal dimension acting on the vacuum; such states are of course simply dual to low energy perturbations of pure AdS~\cite{BanDou98}. The AR puzzle arises from the fact that there are now two apparent bulk geometries -- one with a baby universe and one without -- which are dual to the same CFT state. They propose three possible resolutions: (1) ensemble averaging; (2) invalidity of AdS/CFT: that CFT is not, in fact, dual to AdS;\footnote{In v3 of~\cite{AntRat24}, this was changed to ``beyond AdS/CFT'', suggesting that the CFT would need to be supplemented with extra information. This essentially reduces to the statement that, on its own, the CFT is not dual to a unique bulk.} (3) that the baby universe is not semiclassical. Another possible resolution would be that the two descriptions are somehow ``gauge-equivalent", although it is not exactly clear in what sense such a statement could be made. 

Here we find that (3) is the correct option. Using \textit{only} the extrapolate dictionary in the causal wedge, we construct the aforementioned simple boundary operator $\mathcal{S}_{\partial}$  whose expectation value $\langle {\cal S}_{\partial}\rangle$ conclusively distinguishes between a bulk that has the baby universe 
\begin{equation} \langle {\cal S}_{\partial}\rangle_{\rm if \ bulk \ dual \ has \ baby \ universe}\approx e^{-\log \dim {\cal H}_{\rm baby \ universe}}\end{equation} and a bulk that does not \begin{equation}\langle {\cal S}_{\partial}\rangle_{\rm if \ bulk \ dual \ has \ no \ baby\ universe}=1.\end{equation} We compute the expectation value of this operator and find that it agrees exactly with the case in which the baby universe is absent. The only way in which it can \textit{also} agree with the path integral preparation of the state that includes the baby universe is if the latter has an effective Hilbert space of dimension one. This provides a new argument that uses only the oldest and most conventional aspect of the AdS/CFT correspondence to conclude that the baby universe must have a one-dimensional Hilbert space. Any modification of AdS/CFT which would include a larger Hilbert space for the baby universe would need to give up on the extrapolate dictionary. 

This argument can be extended to a more general setup than the precise AR construction: it clearly applies to any spacetime with a baby universe in which the fine-grained entropy is bounded from above by ${\cal O}(1)$ and which can be approximately truncated to a microcanonical window of energy ${\cal O}(1)$. To generalize even further to states with energy that scales with $N$, we consider states that have (1) no nontrivial QESs in the bulk that are homologous to the complete asymptotic boundary\footnote{A sufficient (but not necessary) condition that guarantees this from the CFT side is that the simple entropy~\cite{EngWal17b, EngWal18} is ${\cal O}(1)$.} and (2) can be represented as an ${\cal O}(1)$ number of local but possibly heavy operators acting on a state of ${\cal O}(1)$ energy.
We show that any such CFT state with a putative baby universe in its entanglement wedge can effectively be reduced to the AR setup: thus in all such constructions, which include AR as a special case, the baby universe is not semiclassically emergent from the CFT. 

To summarize: 
\begin{enumerate} 
    \item We give a concrete proof that the semiclassical baby universe in AR is inconsistent with the extrapolate dictionary. We do this by showing that there must exist operators with duals localized \textit{in the causal wedge} whose expectation values are inconsistent with the existence of more than one semiclassical baby universe state. In particular we explicitly exhibit one such operator ${\cal S}_{\partial}$. This also manifestly excludes the possibility of equivalence of semiclassical descriptions 1 and 2.
    \item We prove that the baby universe can only be consistent with the extrapolate dictionary if it has a one-dimensional Hilbert space. 
    \item We give a test for any putative supplementation of AdS/CFT to include a semiclassical baby universe: it would have to correctly match the operator ${\cal S}_{\partial}$. 
    \item We generalize all of these arguments to arbitrary states with the structure described above. 
    \end{enumerate}

The paper is structured as follows. In Sec.~\ref{sec:assumptions} we state some global assumptions and conventions; assumptions that are necessary in only part of the paper will be stated when needed. Sec.~\ref{sec:ARPuzzle} reviews in the AR puzzle in a convenient framework. Sec.~\ref{sec:swap} constructs the operator ${\cal S}_{\partial}$ and proves that the AR baby universe cannot be semiclassical. We extend our argument to more general states with no nontrivial QESs in Sec.~\ref{sec:newsimple} and conclude in Sec.~\ref{sec:disc}.

\subsection{Assumptions and Conventions}\label{sec:assumptions}
We will make only the following assumptions about the AdS/CFT correspondence:
\begin{enumerate}
    \item \textbf{The extrapolate dictionary.} The algebra of boundary operators ${\cal A}_{\partial M}$ obtained as limits (or via the timelike tube theorem) of bulk operators in $M$ is contained in the algebra of operators of the dual CFT in the large-$N$ limit. 
    \item \textbf{Isometric Encoding in the Causal Wedge:} In the simple setting in which there are no event horizons in the bulk, the low energy EFT states and operators are encoded into the dual CFT via a map that is an application of the HKLL map~\cite{HamKab05, HamKab06, HamKab06b} followed by the extrapolate dictionary. We assume that the HKLL map works (up to small corrections in $N$) as a reconstruction map at sufficiently large but finite-$N$, as explained in~\cite{AlmDon14}; in particular, that the HKLL map is approximately identical at large but finite $N$ on bulks that limit to the same complete causal wedge large-$N$ geometry. In discretizations of the bulk (e.g. tensor networks) the algebra of any subregion is type I, and the encoding map $W:{\cal H}_{\rm bulk}\rightarrow {\cal H}_{\rm CFT}$ is an isometry~\cite{Har16}. In the continuum limit, the encoding map is asymptotically isometric~\cite{FauLi22}: its action on states $\ket{\phi}$ and $\ket{\psi}$ in the code subspace of low energy perturbations limits to an isometry in the infinite-$N$ limit:    
    \begin{equation}
        \lim\limits_{N\rightarrow \infty}\bra{\psi}W_{N}^{\dagger}W_{N}\ket{\phi}=\braket{\psi|\phi}.
    \end{equation}
    A similar statement holds for operators in the causal wedge. We refer readers looking for additional details on asymptotic encoding to~\cite{FauLi22,Ges23}. The encoding map $W$ without a subscript $N$ is the limiting operator of the sequence of operators described above.         
\end{enumerate}
Other assumptions that are not required throughout the paper will only be stated when needed.

\section{The Antonini-Rath Puzzle}\label{sec:ARPuzzle}
\begin{figure}
    \centering
    \includegraphics[width=0.7\linewidth]{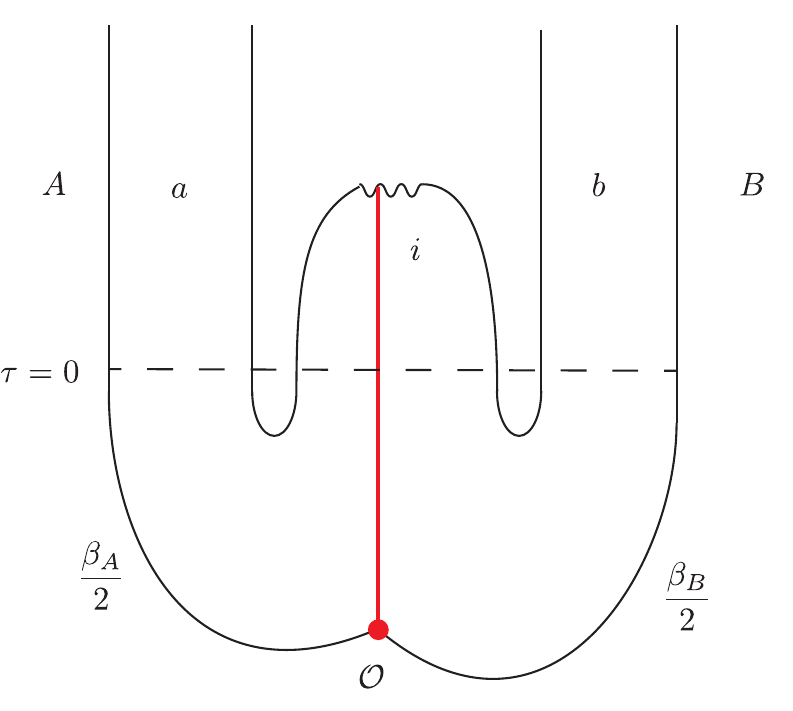}
    \caption{Ref.~\cite{AntSas23}'s path integral preparation of the baby universe, as used by AR in their presentation of the puzzle. There are two asymptotic boundaries $A$ and $B$ at different temperatures, both below Hawking-Page, and a heavy operator insertion ${\cal O}$ in Euclidean time. In Lorentzian time there are three disconnected bulk regions: $a,i,b$.  }
    \label{fig:AntoniniRath}
\end{figure}

We now review the Antonini-Rath (AR) puzzle in language that will be convenient for our subsequent argument. AR consider the setup of~\cite{AntSas23}, which takes two CFTs in a thermal state at a temperature below the Hawking-Page transition and inserts a heavy operator in Euclidean time:
\begin{equation}\label{eq:ARthermal}
\ket{\Psi}_{AB}=\frac{1}{\sqrt{Z}}\sum\limits_{m,n}e^{-\frac{1}{2}(\beta_{A}E_{n}+\beta_{B}E_{m})}O_{m,n}\ket{E_{n}}\ket{E_{m}}.
\end{equation}
To maintain control over tails and fringe effects, AR truncate to a large but ${\cal O}(1)$ microcanonical energy window; we will comment on any possible errors that this may introduce in the next section. At $\beta_{A},\beta_{B}>\beta_{Hawking-Page}$ in this microcanonical window where black holes are very atypical, the entropy of $\Psi_{A}$ (and correspondingly of $\Psi_{B}$) is  large but ${\cal O}(1)$.\footnote{This is the low temperature version of the partially entangled state construction of~\cite{GoeLam18}.} The Euclidean preparation and Lorentzian continuation are illustrated in Fig.~\ref{fig:AntoniniRath}.

Let $\ket{\psi^{(1)}}_{aib}$ be the state of the bulk quantum fields prepared via the Euclidean path integral as in Fig.~\ref{fig:AntoniniRath}. By construction:
\begin{equation}
    S_{\rm vN}[\psi_{i}^{(1)}]= S_{\rm vN}[\psi_{ab}^{(1)}]= \alpha ,
\end{equation}
for some large $\alpha$ that does not scale with $N$. 

AR now consider a rewriting of the CFT state $\ket{\Psi}_{AB}$ in a different basis. For simplicity they consider a bulk theory with a single scalar $\phi$ whose boundary dual is just a single trace operator $\Phi$. Via the state-operator correspondence in the CFT, AR write down a decomposition of $\ket{\Psi}_{AB}$ in terms of local operators of low conformal dimension acting on the vacuum:
\begin{equation}\label{eq:HKLLstate}
    \ket{\Psi}_{AB}
= \sum\limits_{A_{i}B_{j}}c_{A_{i}B_{j}}\Phi_{A_{i}}\Phi_{B_{j}}\ket{0}_{A}\ket{0}_{B} \equiv \sum\limits_{A_{i}B_{j}}c_{A_{i}B_{j}}\ket{A_{i}}\ket{B_{j}}.
\end{equation}

Let us now introduce some notation for the holographic encoding map. Define:
\begin{align}\nonumber
    & W_{a}: {\cal H}_{a}\rightarrow {\cal H}_{A}
\\  
& W_{b}:{\cal H}_{b}\rightarrow {\cal H}_{B}\\ \nonumber
& V:{\cal H}_{ab}\rightarrow {\cal H}_{AB}\end{align}
so that $V=W_{a}\otimes W_{b}$. These are the holographic encoding maps from $a$, $b$, and $ab$ into $A$, $B$, and $AB$. These in particular are the limiting operators in a sequence of operators in $N$ that, when acting on states in the code subspace, asymptotically preserve inner products, expectation values, etc. 

Recall now that as initially shown in~\cite{BanDou98}, there is a bijection between the states $\ket{A_{i}}$, $\ket{B_{j}}$ obtained by acting with the primaries and descendants of the single trace operator $\Phi$ and the states obtained by acting with our local bulk quantum field $\phi$ on the AdS vacuum. AR use this to rewrite:

\begin{align}
    \ket{\Psi}_{AB} &=\sum\limits_{i,j} c_{A_{i}B_{j}} W_{a}(\ket{a_{i}(A_{i})})W_{b}(\ket{b_{j}(B_{j})})\\
    & = V\left (\sum\limits_{i,j} c_{A_{i}B_{j}} \ket{a_{i}}\ket{b_{j}}\right ).
\end{align}
This shows that under the standard holographic encoding map, $\ket{\Psi}_{AB}$ is dual to a single pure state on $ab$ which we shall call $\psi^{(2)}_{ab}$, in apparent contradiction with the fact that it was obtained by Euclidean path integral construction from a bulk that has a mixed state (with large but ${\cal O}(1)$ entropy) on $ab$. AR float several possibilities as possible explanations of this apparent puzzle:\footnote{It is tempting to suggest that there is some problem with the path integral preparation itself, but~\cite{GoeLam18} showed that such states above the Hawking-Page transition are completely conventional and reasonable states in AdS/CFT. It is difficult to see why lowering the temperature to be just below the Hawking-Page transition should all of a sudden make the state illegal.} 
\begin{enumerate}
    \item Ensemble averaging. We will not consider this option here as it cannot resolve the tension in higher dimensions (and it is not clear that it can resolve the tension even in 1+1 dimensions). 
    \item AdS $\neq $ CFT, or equivalently, additional information is needed to determine which asymptotically AdS spacetime is semiclassically emergent from the CFT. 
    \item $i$ is not semiclassical.\footnote{By semiclassical, we would mean that bulk effective field theory is valid up to small corrections in $G$.} The extent of the semiclassical spacetime that is described by the CFT is just the causal wedge $a\cup b$. For example, this is the case if the emergent baby universe Hilbert space is one-dimensional. 
\end{enumerate}
We now give an argument that there is a definitive dual to the CFT, and that the baby universe can only be part of that dual if it has a trivial Hilbert space. We will do this by finding a concrete operator \textit{localized in the causal wedge} that diagnoses the presence of the baby universe. Any putative supplementation of AdS/CFT to include a semiclassical baby universe must pass the test of correctly computing the expectation value of this operator. Once we give the argument in the AR setup, we will generalize it to a larger class of baby universes. 

\section{Swapping Causal Wedges}\label{sec:swap}
Our main tool will be the bulk swap operator.\footnote{Not to be confused with the boundary swap operator, which was studied in the holographic context in~\cite{EngFol24b}.} Let us briefly review the definition of this operator. In a quantum system with Hilbert space ${\cal H}$, the swap operator ${\cal S}$ is a map on pairs of states:
\begin{align*}
    {\cal S}:  & \ {\cal H}\otimes {\cal H}\rightarrow  {\cal H}\otimes {\cal H}\\
    & \ket{\psi}\otimes \ket{\phi}\rightarrow \ket{\phi}\otimes \ket{\psi}.
\end{align*}
Similarly, when applied to density matrices on ${\cal B}({\cal H})$, ${\cal S}$ maps $\rho\otimes \sigma$ to $\sigma \otimes \rho$. Since in the AR setup none of the bulk quantities scale with $N$, this is an ${\cal O}(1)$ complexity operator in the bulk effective field theory, whose expectation value can used as a distinguisher between states. That is, if $\rho$ and $\sigma$ are density matrices, then 
\begin{equation}
    \langle{\cal S}\rangle = {\rm tr}\left [\rho \sigma \right ].
\end{equation}
So, in particular, if $\rho=\sigma$ are identical states $ \langle{\cal S}\rangle ={\rm tr}[\rho^{2}]$. 

To apply ${\cal S}$ to our setup, we consider two copies of the $ab$ system, which we shall label as $ab$ and $a'b'$. We shall similarly consider two copies of our boundary system $AB$, which we shall label as $AB$ and $A'B'$. 
\begin{figure}
\centering
\includegraphics[scale=0.4]{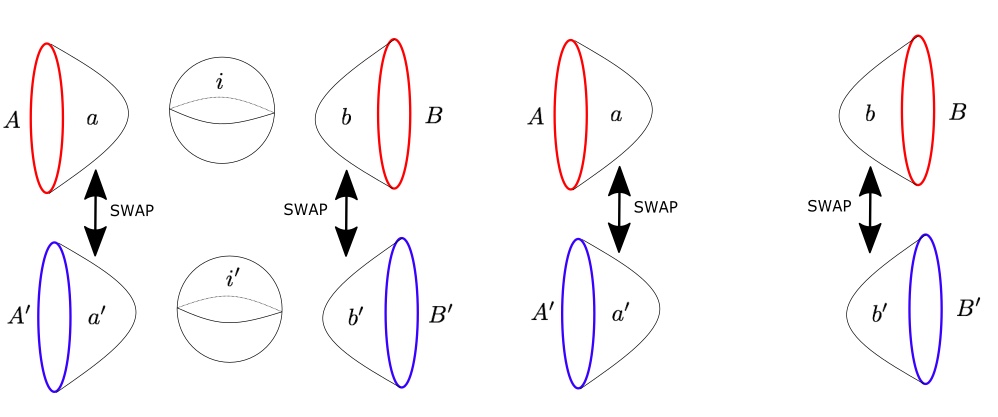}
\caption{The action of the swap operator ${ \cal S}$ on the semiclassical states $\psi^{(1)}\otimes \psi^{(1)}$ (left panel) and $\psi^{(2)}\otimes \psi^{(2)}$ (right panel). This figure represents a time slice of the doubled semiclassical geometries. In both cases the original boundary system is the red system, and the doubled system is the blue system. On the left panel, ${\cal S}$ swaps $ab$ with $a^\prime b^\prime$, but \textit{not} $i$ with $i^\prime$. On the right panel, there is no closed universe so ${\cal S}$ swaps the whole bulk $ab$ with the whole bulk $a^\prime b^\prime$.}
\label{fig:swap}
\end{figure}

We define ${\cal S}$ to be the swap operator that swaps  $ab$ with $a'b'$: thus ${\cal S}$ has support \textit{exclusively} in the causal wedge of $ABA'B'$. It has no support on the baby universe. The action of ${\cal S}$ on $\psi^{(1)}\otimes \psi^{(1)}$ and $\psi^{(2)}\otimes \psi^{(2)}$ is represented on Figure \ref{fig:swap}. We can compute the expectation value of ${\cal S}$ in both $\psi^{(1)}_{ab}$ and $\psi^{(2)}_{ab}$:\footnote{Restricting the action of $\mathcal{S}$ to $ab$ does not change its expectation value as it only acts nontrivially on the causal wedge.}
\begin{align}
    \langle {\cal S}\rangle_{\psi^{(1)}_{ab}\otimes \psi^{(1)}_{a'b'}}&={\rm tr}\left[ \left (\psi^{(1)}_{ab}\right)^{2}\right]\sim e^{-S[\psi_{i}^{(1)}]}\\
    \langle {\cal S}\rangle_{\psi^{(2)}_{ab}\otimes \psi^{(2)}_{a'b'}}&={\rm tr}\left[ \left (\psi^{(2)}_{ab}\right)^{2}\right]=1,
\end{align}
where in the first equation we have used the fact that $\psi_{ab}^{(1)}$ is highly entangled with $i$ over the code subspace. 

Thus we find that \textit{there exists an ${\cal O}(1)$ complexity bulk EFT operator localized purely in the causal wedge, which can distinguish between $\psi_{ab}^{(1)}$ and $\psi_{ab}^{(2)}$.} One may attempt to argue that perhaps ${\cal S}$ breaks entanglement in a way that results in large energy gradients and that it thus cannot be an EFT operator; however, there is only at most ${\cal O}(1)$ entanglement between $ab$ and $i$, and this entanglement is not UV entanglement: we are only ever swapping complete disconnected universes. 

Since ${\cal S}$ is localized to the causal wedge $aba'b'$, there is no ambiguity on whether it admits a boundary dual or not: it does. In addition, there is no ambiguity about how to represent ${\cal S}$ on the boundary: since ${\cal S}$ is a causal wedge operator of ${\cal O}(1)$ complexity, it can be represented on the boundary using the standard causal wedge encoding map. Here we leave open the possibility that the encoding map is somehow unusual or different for the baby universe: because ${\cal S}$ is a causal wedge operator, that possibility is irrelevant to our argument; this is why in Sec.~\ref{sec:assumptions} our assumptions about the encoding map were exclusively about its action on the causal wedge. The encoding map $V\otimes V$ yields a representation ${\cal S}_{\partial}$ of ${\cal S}$ in the CFT on $ABA'B'$ such that \footnote{Note that the resulting boundary operator is \textit{not} identical to the boundary swap operator.}
\begin{equation}
    {\cal S}_{\partial}(V\otimes V)= (V\otimes V){\cal S}.
\end{equation}
We now compute the expectation value of ${\cal S}_{\partial}$ in $\ket{\Psi}_{AB}$. Because $V$ is an isometry, if $\langle {\cal S}_{\partial}\rangle =1$, $\ket{\Psi}_{AB}$ must be dual to description 2, whereas if $\langle {\cal S}_{\partial}\rangle \ll 1$, the dual must be description 1.

Recall now that during our review of AR's work, we established that
\begin{equation}
    \Psi_{AB}= V \psi^{(2)}_{ab} V^{\dagger}.
\end{equation}
Thus
\begin{equation}
    \langle {\cal S}_{\partial}\rangle_{\Psi_{AB}\otimes \Psi_{A'B'} }= \langle {\cal S}_{\partial}\rangle_{V \psi^{(2)}_{ab} V^{\dagger}\otimes V\psi_{a'b'}^{(2)}V^{\dagger}}=\langle {\cal S}\rangle _{\psi^{(2)}_{ab}\otimes\psi^{(2)}_{a'b'}}=1,
    \label{eq:hong}
\end{equation}
where we have used the fact that the encoding of the causal wedge is isometric, and isometries preserve expectation values. We remind the reader that ${\cal S}_{\partial}$ is \textit{not} the boundary swap operator, so Eq.~\ref{eq:hong} does not just follow from purity of $\ket{\Psi}_{AB}$ but rather from the purity of $\psi^{(2)}_{ab}$.

There is thus \textit{no} ambiguity in the dual of $\ket{\Psi}_{AB}$. We found a boundary operator ${\cal S}_{\partial}$ whose dual operator ${\cal S}$ is localized in the causal wedge, and this  ${\cal S}$ definitively distinguishes between the bulk state with the baby universe and the bulk state without it. We have computed the expectation value of the corresponding CFT operator ${\cal S}_{\partial}$ and found that it conclusively agrees with the state in which the baby universe is absent. 

We thus find that AdS/CFT has a definitive answer about the dual, assuming (a) the extrapolate dictionary, and (b) isometricity of the encoding map localized to the causal wedge in the absence of horizons. These two aspects of AdS/CFT are arguably the most well-established aspects of the duality.

How, then, are we to understand $\psi_{aib}^{(1)}$ and consistency of the gravitational path integral picture? If we insist on validity of the path integral approach, we must have 
\begin{equation}
    e^{-S[\psi_{i}^{(1)}]}\sim\langle {\cal S}\rangle_{\psi^{(1)}_{ab}\otimes \psi^{(1)}_{a'b'}}
    =\langle {\cal S}\rangle_{\psi^{(2)}_{ab}\otimes \psi^{(2)}_{a'b'}}=1,
\end{equation}
which is only possible if $|{\cal H}_{i}|=1$. If we were to apply an encoding map to $aib$, the only way in which we could obtain an answer that is consistent with $\langle {\cal S}_{\partial}\rangle=1$ is if the baby universe is not semiclassical: that is, the holographic encoding map must throw out the baby with the bathwater. In particular, our results show that any attempts to create an emergent semiclassical baby universe by modifying or altering AdS/CFT would need to find a way of matching the expectation value of ${\cal S}_{\partial}$ without destroying the extrapolate dictionary or the isometric encoding map of the causal wedge.

It is tempting to try to find a loophole in this argument by recalling that the state in Eq.~\ref{eq:ARthermal} required a truncation of the tails to a microcanonical window. This suggests a potential way out: that in computing $\langle {\cal S}\rangle_{\psi_{ab}^{(1)}}$ from the boundary, we would find the correct answer of $e^{-S_{\rm vN}[\psi_{i}^{(1)}]}$ if we had just included the tail. However, this cannot be the case: recall the exact expression of the finite $N$ state prepared by the gravitational path integral in Eq.~\ref{eq:ARthermal}:
\begin{align}\ket{\Psi}_{AB}=\frac{1}{\sqrt{Z}}\sum\limits_{m,n}e^{-\frac{1}{2}(\beta_{A}E_{n}+\beta_{B}E_{m})}O_{m,n}\ket{E_{n}}\ket{E_{m}}.\end{align}The energy $\mathcal{E}=E_L+E_R$ of the two-sided state is then described by a random variable of expectation value $E_0=\mathcal{O}(1)$. Let us choose $\Delta_0\gg E_0$ but still $\mathcal{O}(1)$. By Markov's inequality, \begin{align}\mathbb{P}(\mathcal{E}>\Delta_0)\leq \frac{E_0}{\Delta_0}.\end{align} This means that \begin{align}\frac{1}{Z}\sum_{E_n+E_m>\Delta_0}e^{-(\beta_AE_n+\beta_BE_m)}\vert O_{m,n}\vert^2\leq\frac{E_0}{\Delta_0},\end{align}and so we deduce \begin{align}\frac{1}{Z}\sum_{E_n>\Delta_0\,\mathrm{or}\, E_m>\Delta_0}e^{-(\beta_AE_n+\beta_BE_m)}\vert O_{m,n}\vert^2\leq\frac{E_0}{\Delta_0}.\end{align}If we denote, like in \cite{AntRat24}, by $P_{\Delta_0,(L,R)}$ the projections onto the microcanonical windows of size $\Delta_0$ on the left and on the right respectively, we obtain that \begin{align}\|\ket{\Psi_{AB}}-P_{\Delta_0,L}P_{\Delta_0,R}\ket{\Psi_{AB}}\|^2\leq \frac{E_0}{\Delta_0},\end{align} and can therefore be made arbitrarily small since we are free to choose $\Delta_0$ as large as we want as long as it is $\mathcal{O}(1)$. Now, by the triangle inequality and the Cauchy--Schwarz inequality, if $X$ is a two-sided boundary observable,
\begin{align}\lvert\mathrm{tr}((\Psi_{AB}-P_{\Delta_0,L}P_{\Delta_0,R}\Psi_{AB}P_{\Delta_0,L}P_{\Delta_0,R})X)\rvert&\leq2\|\ket{\Psi_{AB}}-P_{\Delta_0,L}P_{\Delta_0,R}\ket{\Psi_{AB}}\|\|X\|\\&\leq2\sqrt{\frac{E_0}{\Delta_0}}\|X\|.\end{align}This shows that for $\Delta_0$ chosen large enough (but $\mathcal{O}(1)$),\begin{align}\|\Psi_{AB}-P_{\Delta_0,L}P_{\Delta_0,R}\Psi_{AB}P_{\Delta_0,L}P_{\Delta_0,R}\|_1\ll 1.\end{align}
Thus the untruncated and truncated states are close in 1-norm. Since reconstruction of the causal wedge is isometric, and the 1-norm can't grow on the preimage of an isometry, this means that if the states corresponding to the untruncated and truncated boundary states are $\psi_{ab}^{(1)}$ and $\psi_{ab}^{(2)}$, they must also be close in 1-norm:
\begin{equation}
    ||\psi_{ab}^{(1)}-\psi_{ab}^{(2)}||_{1}\ll 1.
\end{equation}
We now prove that this is not possible given the expectation value of the swap operator in the two different states. By Holder's Inequality:
\begin{equation}
    |{\rm tr}\left [\left (\rho-\sigma\right){\cal S}]\right | \leq ||\rho-\sigma||_{1}||{\cal S}||_{\infty},
\end{equation}
where $||\cdot ||_{\infty}$ refers to the operator norm, which is 1 for ${\cal S}$, $\rho=\psi_{ab}^{(1)}\otimes\psi_{ab}^{(1)}$, and $\sigma=\psi_{ab}^{(2)}\otimes\psi_{ab}^{(2)}$. This shows:
\begin{equation}
    |\langle {\cal S}\rangle_{\psi_{ab}^{(1)}\otimes\psi_{ab}^{(1)}}-\langle {\cal S}\rangle_{\psi_{ab}^{(2)}\otimes\psi_{ab}^{(2)}} | \leq ||\rho-\sigma||_{1}.
\end{equation}
We thus find: 
\begin{equation}
||\psi_{ab}^{(1)}\otimes\psi_{ab}^{(1)}-\psi_{ab}^{(2)}\otimes\psi_{ab}^{(2)}||_{1}\gtrsim 1-e^{-S_{\rm vN}[\psi_{i}^{(1)}]}. 
\end{equation}
The triangle inequality, together with the fact that the 1-norm is a cross norm, then yields the necessary result:
\begin{equation}||\psi_{ab}^{(1)}-\psi_{ab}^{(2)}||_{1}\geq \frac{1}{2}||\psi_{ab}^{(1)}\otimes\psi_{ab}^{(1)}-\psi_{ab}^{(2)}\otimes\psi_{ab}^{(2)}||_{1}\gtrsim \frac{1}{2}(1-e^{-S_{\rm vN}[\psi_{i}^{(1)}]}).
\end{equation}

Therefore it is not possible that any small differences between the state in Eq.~\ref{eq:ARthermal} prior to truncation to the microcanonical window and the state in Eq.~\ref{eq:HKLLstate} could account for the large discrepancy in the expectation values of the swap operator. For both $\psi^{(1)}$ and $\psi^{(2)}$ to have CFT duals, there are only two options: (1) the duals are nearly orthogonal, or (2) the entropy of the baby universe is bounded from above by 0. Since the duals are the same state up to a truncation of the thermal tails, we conclude option (2): if it is to be encoded into the CFT, the baby universe cannot be semiclassical.

We note here that while our operator $\langle {\cal S}_{\partial}\rangle$ acts on two copies of the state $\ket{\Psi}_{AB}\otimes \ket{\Psi}_{A'B'}$, it must be the case that there exists another operator with the same ability to distinguish $\psi^{(1)}_{ab}$ from $\psi^{(2)}_{ab}$ but which only acts on a single copy. To see this, we use Fannes' Inequality:
\begin{equation}
    |S(\psi^{(1)}_{ab})-S(\psi^{(2)}_{ab})|\leq \frac{1}{2}||\psi^{(1)}_{ab}-\psi^{(2)}_{ab}||_{1} \log\Delta_{0} + \frac{1}{e}
\end{equation}
where $\Delta_{0}$ sets the dimension of the code subspace. Since $\psi^{(1)}_{ab}$ is by construction approximately maximally entangled within this code subspace, $S(\psi^{(1)}_{ab})\sim \log \Delta_{0}$. We thus find that \begin{equation}
||\psi^{(1)}_{ab}-\psi^{(2)}_{ab}||_{1}\geq 2-{\cal O}(1/\log\Delta_{0}),
\end{equation}
Recall now that:
\begin{equation}\label{eq:finalFannes}
    ||\psi^{(1)}_{ab}-\psi^{(2)}_{ab}||_{1} = \sup\limits_{\|X\|=1}\left\lvert{\rm tr}\left [ X(\psi_{ab}^{(1)}-\psi_{ab}^{(2)})\right]\right\rvert.
\end{equation}
Thus we find that there is an operator that can definitively distinguish between $\psi_{ab}^{(1)}$ and $\psi_{ab}^{(2)}$ even given just one copy of the system. Since this operator is localized to $ab$ -- i.e., the causal wedge -- it can be encoded into the CFT via an isometry, which will preserve expectation values. Thus there exists an operator on $AB$ that can conclusively ascertain if the state $\ket{\Psi}_{AB}$ is dual to $\psi_{aib}^{(1)}$ or $\psi_{ab}^{(2)}$. This does not, however, provide a construction of such an operator, and one could worry that this operator somehow destroys the EFT. Moreover, finding an explicit causal wedge operator whose expectation values drastically differ in the two descriptions is valuable in the sense that one can then precisely ask how to restore a consistent expectation value for this operator in the presence of a semiclassical baby universe. We will further comment on this in Section \ref{sec:disc}.  For this reason we have opted to present our arguments in terms of the explicit operator ${\cal S}_{\partial}$. Let us briefly note an additional consequence of Eq.~\ref{eq:finalFannes}: there is a distinguishing operator if $S_{\rm vN}[\psi_{ab}^{(1)}]>0$ can be made to scale like $\mathrm{log}\;\Delta_0$. On the other hand, if the baby universe Hilbert space is one-dimensional, there is no distinguisher.

This concludes our argument for the baby universe in the AR construction. We now proceed to a more general picture that does not rely on a path integral construction. 

\section{Baby Universes Simply Can't Result from Simple Black Holes}\label{sec:newsimple}

Our argument in the previous section relied very essentially on the absence of a QES homologous to the complete asymptotic boundary with generalized entropy larger than ${\cal O}(G^{0})$. It turns out that at the cost of one additional standard AdS/CFT assumption, we obtain the same result in any putative AdS/CFT baby universe construction in which there is no QES with generalized entropy larger than ${\cal O}(G^{0})$. We now provide the argument, starting with the following assumption:

\vspace{0.1cm}

\noindent \textbf{Assumption:} the simple entropy construction~\cite{EngWal17a, EngWal18, EngPen21a}, which iteratively removes all sources that result in non-stationarity of horizons, applies when the horizon changes topology to the empty set. For readers unfamiliar with the simple entropy, we now briefly review this result. The primary observation is that horizons are only non-stationary due to matter (or gravitational radiation) falling across them from the asymptotic region. Thus if we can turn off the appropriate sources at $\mathscr{I}$, we would remove the infalling or outgoing matter, resulting in a stationary horizon. In a spacetime with no nontrivial QESs homologous to the complete asymptotic boundary, this removes horizons altogether. The work of~\cite{EngPen21a} proved this in the absence of such topological transitions, but asserted that the result should continue to hold under such topological transitions. We will operate under the assumption that it does indeed continue to hold. This is certainly true in e.g. Vaidya thin-shell collapse: turning off the heavy operator sourcing the shell restores the spacetime to pure AdS. We could also act on the AR state with heavy operators in Lorentzian time; that would also satisfy this assumption. 
 
\vspace{0.1cm}

Note that we did not need to make this assumption in the AR setup as we had started with a horizonless geometry and an explicit state. 

We now proceed to state the argument. Consider a bulk spacetime with the following properties:
\begin{enumerate}
    \item A connected asymptotically AdS piece, which we shall call $a$. We are agnostic about how many asymptotic boundaries $a$ has. (In the previous section, we denoted each complete connected bulk region with its own letter, e.g. $a$, $b$ etc; here to maintain full generality we will use $a$ to refer to all regions with an asymptotic boundary.)
    \item A disconnected, boundaryless baby universe, which we shall call $i$. 
    \item No nonempty QESs that are homologous to the asymptotic boundary(ies) $A$. 
\end{enumerate}

Let $\ket{\Psi}_{A}$ be a CFT state dual to this geometry with some bulk state on it. We make one further assumption, which is that $\ket{\Psi}_{A}$ can be represented as an ${\cal O}(1)$ number of simple (but possibly heavy) operators acting on a state of ${\cal O}(1)$ energy. This is always true for a black hole formed from fast collapse. Even though the minimal QES is always the empty set, the entanglement wedge of $\ket{\Psi}_{A}$ depends on the entanglement of the bulk state: if the bulk state has  entanglement between $a$ and $i$, the entanglement wedge of the pure state $\ket{\Psi}_{A}$ includes $i$. Otherwise, we will argue that it does not.

\begin{figure}
    \centering
    \includegraphics[width=0.8\linewidth]{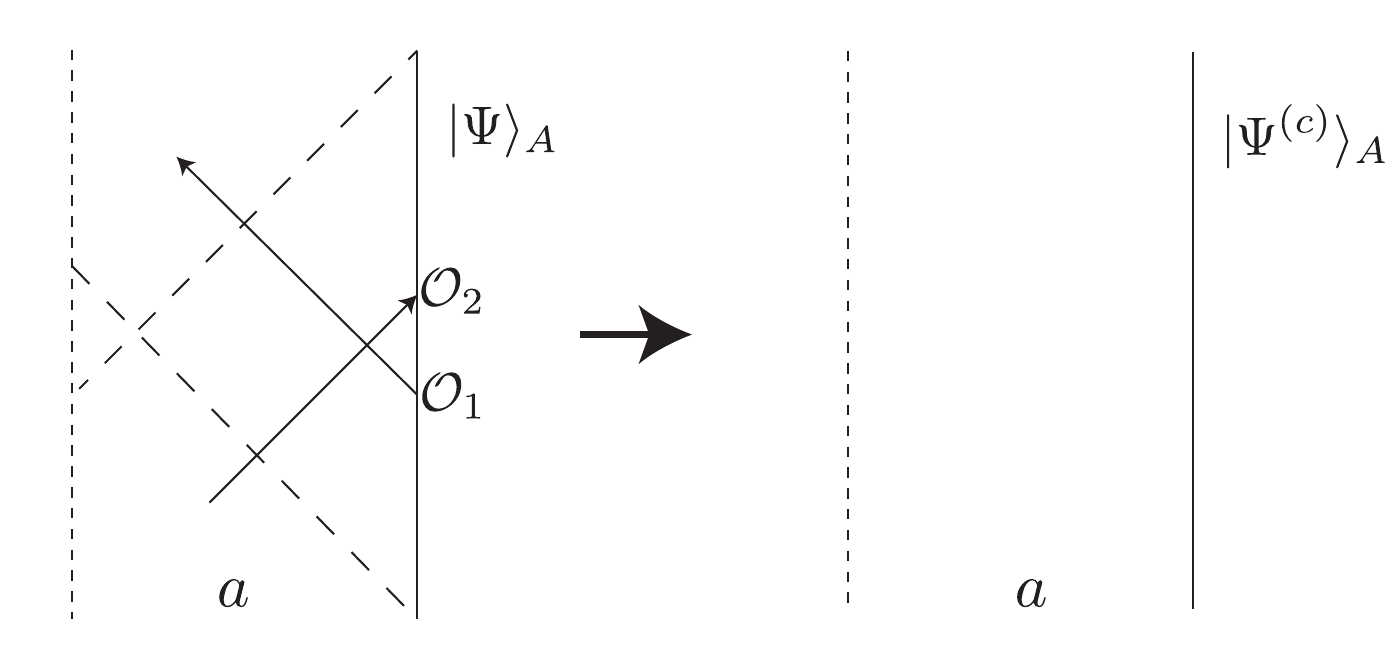}
    \caption{The state $\ket{\Psi}_{A}={\cal O}_{1}{\cal O}_{2}\ket{\Psi^{(c)}}_{A}$. Turning off the sources removes the event horizon.}
    \label{fig:sources}
\end{figure}

If $a$ has event horizons, we will remove them using the protocol of~\cite{EngPen21a}: we turn on sources that (on some timefold) propagate locally from the AdS boundary and push any event horizons to the outermost QES. Since there are no nonempty QESs, this protocol removes all horizons from $a$.\footnote{In the language of \cite{GesLiu24}, this turns a state with an infinite causal depth parameter into a state with finite causal depth parameter.} We will  denote by $\ket{\Psi^{(c)}}_{A}$ the CFT state that has been acted on in this way to reveal all of the AdS region as a causal wedge. See Fig.~\ref{fig:sources}. We emphasize that the sources involved propagate causally from the boundary: they cannot act on the baby universe or change its state. 

By our assumption on $\ket{\Psi}_{A}$, $\ket{\Psi^{(c)}}_{A}$ has ${\cal O}(1)$ energy.  Now we use our proof from Sec.~\ref{sec:swap} that the truncation of a state with ${\cal O}(1)$ energy to a microcanonical window of size ${\cal O}(1)$ cannot result in a new state that has a large trace distance from the untruncated state, at least in the infinite-$N$ limit. This means that $\ket{\Psi^{(c)}}_{A}$ is well approximated by a state in microcanonical window of ${\cal O}(1)$, which we shall  call $\ket{\Psi^{(c)'}}_{A}$.

Thus far our discussion has been independent of the exact state of the bulk. We now consider a low energy bulk state $\ket{\psi}_{ai}$, where there is entanglement between the AdS region and the baby universe. 

Let us now introduce a second copy of $a$, which we shall term $a'$. We define, as in the previous section, the bulk swap operator ${\cal S}$ to swap bulk states on $a$ and $a'$. It immediately follows from Sec.~\ref{sec:swap} that
\begin{align}
    & \langle {\cal S}\rangle_{\psi^{(1)}_{a}\otimes \psi^{(1)}_{a'}}\sim e^{-S_{\rm vN}[\psi_{a}^{(1)}]}.
\end{align}

We recall that a CFT state at large-$N$ in an ${\cal O}(1)$ microcanonical window admits a decomposition in terms of operators of low conformal dimension acting on the vacuum. Once again invoking~\cite{BanDou98}, the dual of such a state is simply a low energy perturbation of AdS: no baby universe included. The expectation value $\langle {\cal S}_{\partial}\rangle=1$, and we find that the entanglement wedge of $\ket{\Psi^{(c)'}}_{A}$ does not contain the baby universe unless $|{\cal H}_{i}|=1$, and then the baby universe  cannot be semiclassical.

Can $\ket{\Psi^{(c)}}_{A}$ contain the baby universe in its entanglement wedge? The answer is no: because $\ket{\Psi^{(c)'}}_{A}$ and $\ket{\Psi^{(c)}}_{A}$ are close in trace distance, $\langle {\cal S}_{\partial}\rangle$ must be approximately the same for the two states. 

Can the original state prior to turning on simple sources, $\ket{\Psi}_{A}$, contain a semiclassical baby universe in its entanglement wedge? Once again, no: the locally propagating sources in question do not act on the baby universe and do not modify bulk entanglement of the state in the causal wedge. We thus arrive at the same conclusion as we did in the previous section: if there are no nontrivial QESs homologous to $A$, then under the stated assumptions the CFT definitively picks the state without a baby universe -- unless the baby universe is contentless: $|{\cal H}_{i}|=1$, and $i$ is not semiclassical.

\section{Discussion}\label{sec:disc}

In this paper, we constructed a simple boundary operator whose bulk dual has support exclusively in the causal wedge, which can conclusively diagnose if a semiclassical baby universe is encoded in the CFT. In a large class of semiclassical geometries with a closed universe, including the one of \cite{AntRat24}, the verdict is definitive: the closed universe cannot emerge in the AdS/CFT correspondence. Our argument only relies on the most basic features of the holographic dictionary: the extrapolate dictionary and the (asymptotically) isometric encoding of the causal wedge. While it is reassuring that as decades of research have shown, AdS does indeed equal CFT, failure of semiclassicality of closed universes is of course a very embarrassing state of affairs, given that we may well live in a closed cosmology. Even without worrying about our own world, this conclusion makes it very challenging to think about the interior of small black holes in the bulk of AdS at late times from the point of view of the CFT.

Recent developments \cite{AbdSte25,HarUsa25} have suggested that one way to recover semiclassical physics in a closed universe is to explicitly include an observer in the range of the holographic map~\cite{HarUsa25, AbdSte25}. It would be very interesting to understand how the calculation performed in this paper, and in particular the expectation value we found for the swap operator $\mathcal{S}_{\partial}$, is affected by conditioning on the presence of an observer inside the closed universe. It would also be interesting to understand what happens to these arguments in the presence of a Python's Lunch~\cite{BroGha19, EngPen21a, EngPen21b, EngPen23}, which violates the assumptions of Sec.~\ref{sec:newsimple}. We leave further investigations of the implications of the bulk swap operator's encoding on the boundary to future work.
\section*{Acknowledgments}
It is a pleasure to thank D. Harlow, J. Sorce,  and H. Liu for helpful discussions. The work of NE is supported in part by the Department of Energy under Early Career Award DE-SC0021886, by the Heising-Simons Foundation under grant no. 2023-4430,  and by the Templeton Foundation via the Black Hole Initiative. The work of EG is supported in part by grant NSF PHY-2309135 to the Kavli Institute for Theoretical Physics (KITP), as well as the Todd Alworth Larson fellowship.

\bibliographystyle{jhep}
\bibliography{all}

\end{document}